\documentclass[fleqn,10pt]{wlscirep}

\usepackage{booktabs}
\usepackage{multirow}
\usepackage{longtable}

\newcommand*\sfref[1]{%
    Supplementary Figure \ref{#1}}
\newcommand*\stref[1]{%
    Supplementary Table \ref{#1}}
\newcommand*\smref[1]{%
    Supplementary Text \ref{#1}}

\title{Identifying latent activity behaviors and lifestyles using mobility data to describe urban dynamics}

\author[a,b]{Yanni Yang}
\author[b]{Alex Pentland} 
\author[b,c,1]{Esteban Moro}

\affil[a]{Department of Computing, The Hong Kong Polytechnic University, Hong Kong, China}
\affil[b]{Connection Science, Institute for Data Science and Society, MIT, Cambridge, USA}
\affil[c]{Department of Mathematics and GISC, Universidad Carlos III de Madrid, Legan\'es, Spain}

\affil[*]{corresponding author(s): Esteban Moro (E-mail: emoro@mit.edu)}

\begin{abstract}
Urbanization and its problems require an in-depth and comprehensive understanding of urban dynamics, especially the complex and diversified lifestyles in modern cities. Digitally acquired data can accurately capture complex human activity, but it lacks the interpretability of demographic data. In this paper, we study a privacy-enhanced dataset of the mobility visitation patterns of 1.2 million people to 1.1 million places in 11 metro areas in the U.S. to detect the latent mobility behaviors and lifestyles in the largest American cities. Despite the considerable complexity of mobility visitations, we found that lifestyles can be automatically decomposed into only 12 latent interpretable activity behaviors on how people combine shopping, eating, working, or using their free time. Rather than describing individuals with a single lifestyle, we find that city dwellers' behavior is a mixture of those behaviors. Those detected latent activity behaviors are equally present across cities and cannot be fully explained by main demographic features. Finally, we find those latent behaviors are associated with dynamics like experienced income segregation, transportation, or healthy behaviors in cities, even after controlling for demographic features. Our results signal the importance of complementing traditional census data with activity behaviors to understand urban dynamics.
\end{abstract}
\begin{document}

\flushbottom
\maketitle



\section*{Introduction}
Cities are the main ground on which our society and culture develop today. Most of our current understanding of problems like transportation, mobility, inequality, gentrification, or even social participation is based on census or survey information, which is updated infrequently, contains only coarse-grain information and is scattered across different agencies or institutions \cite{Cagney.2020}. On the other hand, we now have the potential to complement official data with high-resolution updates on how people purchase, move, get a job, or interact by leveraging new sources of information from mobile data \cite{Song:2012wk,eagle2009eigenbehaviors}, social media \cite{llorente2014social,huang2016activity}, wifi networks \cite{kontokosta2017urban,bellini2017wi}, phone apps \cite{moro,athey}, and credit cards \cite{di2018sequences,coco2013}. Companies have been using this wealth of data in the past. They are currently able to micro-segment clients based on their demographics and their behavioral traits \cite{esri2018,mitchell1983nine,kahle1986}. However, most cities are still using primary segments of census groups (residential areas, housing prices, gender, age, unemployment) or small behavioral surveys to map problems like inequality, gentrification, or transportation. This approach falls short of anticipating, monitoring, or forecasting the rapid and complex evolution of those problems in our cities. For example, the recent pandemic has highlighted the shortcomings of using outdated, non-integrated, and slow processed data to manage and anticipate the spreading of COVID19 and the special relevance of real-time, more granular, and high-frequency mobility data \cite{nuria2020,aleta2020}.

In particular, people's mobility data has become more available thanks to the prevalence of location acquisition techniques and mobile phones, and it enables a new way to study and understand human behavior in cities. People's mobility behavior, e.g., the places they visit and their visiting frequency, can reflect people's lifestyle, understood as ``the way in which a person or group lives'' \cite{mitchell1983nine,zion2018identifying,kitamura2009}. Given the importance of lifestyles to predict individual and a group of individual's behavior, they have been thoroughly explored mainly in marketing \cite{mitchell1983nine} but also in many fields from transportation, \cite{Salomon.1983}, health \cite{sadilek2013,joumard2010health,matz2015urban} to psychology and sociology \cite{kahle1986,hanson1981travel}. 

The study of activity patterns and detection of lifestyles of urban residents based on survey data has a long tradition, \cite{hanson1981travel} but recent developments in data collection and analysis have allowed the unveiling of the high-dimensional, rapid-changing, and complex lifestyles in our cities \cite{kitamura2009,Gonzalez:2009p1717,zhao2020discovering,Salomon.1983,di2018sequences,hu2016tales,zion2018identifying,Farrahi:2011fka,mostafavi2022}. Studies that try to detect those lifestyles from activity data are generally limited by the completeness of the activity/mobility space (only expenditure patterns \cite{di2018sequences}, mobility patterns only when mobile phone calls and messages appear \cite{xu2019mining,aledavood2022}, only transportation transit patterns \cite{zhao2020discovering}, or a very small number of demographic variables \cite{Salomon.1983}), the limited geography (only one or two cities \cite{di2018sequences, hu2016tales}), or the number of people used to detect lifestyles \cite{zion2018identifying,Farrahi:2011fka}. As a result, a small number of meaningful lifestyles were detected, insufficient to accommodate the highly heterogeneous and complex variability of our cities' behaviors.

On the other hand, in those studies, city residents' behaviors are typically classified into a single lifestyle group \cite{Salomon.1983,di2018sequences}. This forces us to divide very similar individual behaviors into different groups just based on slight differences. Consequently, a significant fraction of individuals end up with unclassified lifestyles groups \cite{di2018sequences}, or across groups with minimal different characteristics \cite{Salomon.1983}. These problems severely limit those lifestyle groups' potential applicability to understanding problems like social-economic integration, mobility, or health, since slight individual behavioral differences or even different or incomplete datasets can yield a different grouping of users or lifestyles \cite{hill2009can,kitamura2009,jiang2012clustering,xi2020beyond}.

In this work, we uncover people's lifestyles using a dataset of mobility traces of more than 1.2 million anonymous, opted-in users in 11 cities in the United States. By formulating people's behavior using venue and temporal activity vectors, we can extract a set of interpretable latent activity behavioral patterns \cite{toch2019}. Those latent behaviors are groups of visitation patterns that frequently co-occur in our sample of users. People's lifestyle is not a label for each individual but rather a linear combination of those latent behaviors with different weights. We investigate whether those behaviors can be predicted by simple demographic traits, e.g., race, income, or transportation. Although we find a small correlation, latent behaviors seem to be primarily independent of those demographic traits. Finally, we find that each component of those latent patterns has a different relationship with social, mobility, and health problems. Our results indicate that it is possible to construct a {\em behavioral rich census} of lifestyles in the U.S. cities that can complement traditional census to understand the main processes and problems in our cities.

\section*{Results}

Our primary data source is from Cuebiq, a location intelligence, and measurement company that, in 2017 supplied six-month-long records of anonymized, privacy-enhanced, and high-resolution mobile location pings across 11 U.S. census core-based statistical areas (CBSAs), see \smref{sectionSM:data}. It consists of approximately 67 billion records from $N = $1.2 million anonymous opted-in devices, each of which has reported a total of at least 2,000 locations over the six-month observation period. Our second data source is a collection of approximately 1.1 million verified venues across all CBSAs, obtained via the Foursquare API in 2017. Those venues are classified into different categories according to the Foursquare Category Hierarchy \cite{catsfsq}. We infer the {\em home area} of each individual at the Census Block Group level using their most common location between 10 pm and 6 am. We further extract any individual visits to a given place that lasts for more than 5 minutes (see \smref{sectionSM:data}) and are less than 4 hours long. It is important to note that our visitation patterns include not only consumption patterns (restaurants, shops, sports events, etc.) but also other non-commercial activities (transportation, education, health, outdoor activities, etc.), which are important to explain urban lifestyles.

\begin{figure*}[bt!]
\centering
\includegraphics[width=0.99\textwidth]{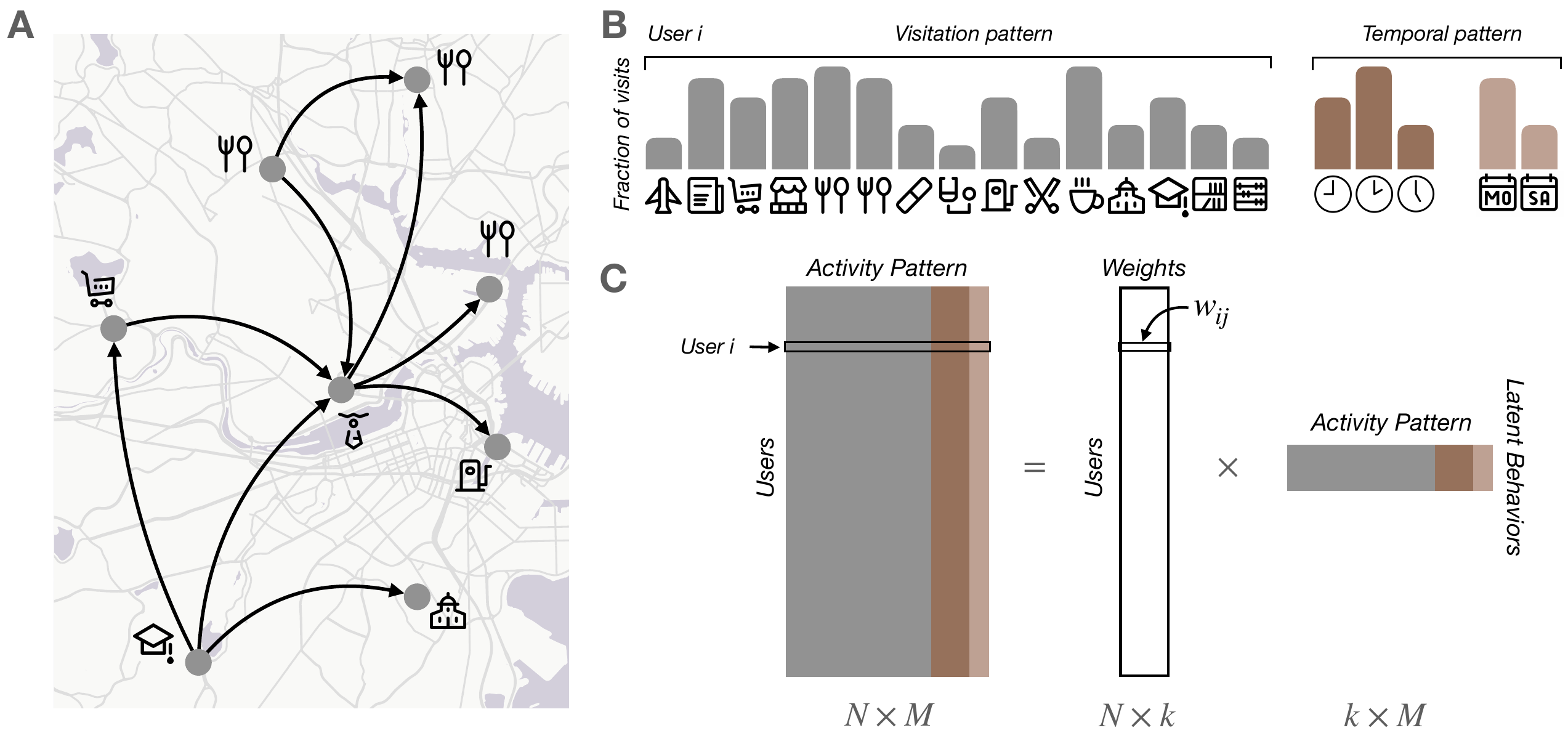}
\caption{{\bf Detecting latent behaviors} A) Using individuals' trajectories, we identify the visits to the different places and the categories of those places. B) Each individual is described by a $M$-dimensional (normalized) vector that contains the fraction of visits to each of the 286 categories (visitation pattern) plus the fraction of visits at different times during the day and the week (temporal pattern). C) Non-negative matrix factorization is used to decompose the matrix of the $M$-dimensional vectors for each of our $N$ users into a matrix of $k$ different latent behaviors and the corresponding behavior's weights for each user. Icons designed by bqlqn/flaticon.com and Boston maps produced using Open Street Map data.}
\label{figure1}
\end{figure*}

Rather than describing a person by a unique pattern, we will assume that there are some latent behavioral patterns that, when combined, define a person's lifestyle. The weight of the different latent behavior patterns could reflect their dominance over the person's lifestyle. To detect those latent behavioral patterns, we describe the activity of each user $i$ by a $M$ dimensional vector, which includes the (normalized) number of visits to the different types of places (248 venue categories, see Figure \ref{figure1} and \sfref{figSM:allcomponents}). We also include five temporal features about the fraction of those visits that happen during the morning (5 am to noon), afternoon (noon to 6 pm), evening/nighttime (6 pm to midnight), and also during the weekend and weekdays. Thus user $i$ activity is described by a vector $\mathbf{x}_i$ of $M = 248+5$ components. When put together for all $N$ users, they form a matrix $X$ of $N \times M$ dimensions. Different methods exist to learn the latent patterns for the vectors of those $N$ users, like spectral methods \cite{eagle2009eigenbehaviors}, Latent Dirichlet Allocation \cite{Farrahi:2011fka,xu2019mining,zhao2020}, neural networks, \cite{zion2018identifying} or complex networks \cite{di2018sequences}. In general, these methods detect latent patterns, i.e. co-occurrence of variables (visitations in our case) that frequently appear in the dataset. Given that our vectors are non-negative, we apply non-negative matrix factorization (NMF) to $X$. NMF is a powerful technique for finding parts-based, linear representations of non-negative data and has been applied successfully in several applications like genomics, image recognition, or text mining \cite{lee99,brunet2004} (see Materials and Methods). In the context of human behavior, it has also been used to identify the activity patterns of users or urban areas \cite{Gauvin:2014fca,GraellsGarrido:tq,hu2016tales,aledavood2022,mollgaard2022}. 
 
The key idea is that the activity matrix can be decomposed into two matrices $X = W \cdot B$, where $B$ is a matrix of dimension $k \times M$ that contains each of the $k$ latent behaviors and $W$ is a $N \times k$ matrix that contains the weights of those latent behaviors for each user (see Figure \ref{figure1}C). Thus each $\mathbf{x}_i$ can be decomposed as linear combination of $k$ latent behaviors $\mathbf{x}_i = \sum_{j=1}^k w_{ij} \mathbf{b}_j$. In each latent behavior pattern $\mathbf{b}_j$, the higher value the type of venue or temporal feature holds, the more dominant that category of place or time of the week acts in the latent behavior pattern. For each user $i$, the higher the $w_{ij}$, the more important is latent behavior $\mathbf{b}_j$ to explain her activity.

\subsection*{Detection of latent behaviors}
\begin{figure*}[!ht]
\centering
\includegraphics[width=0.99\textwidth]{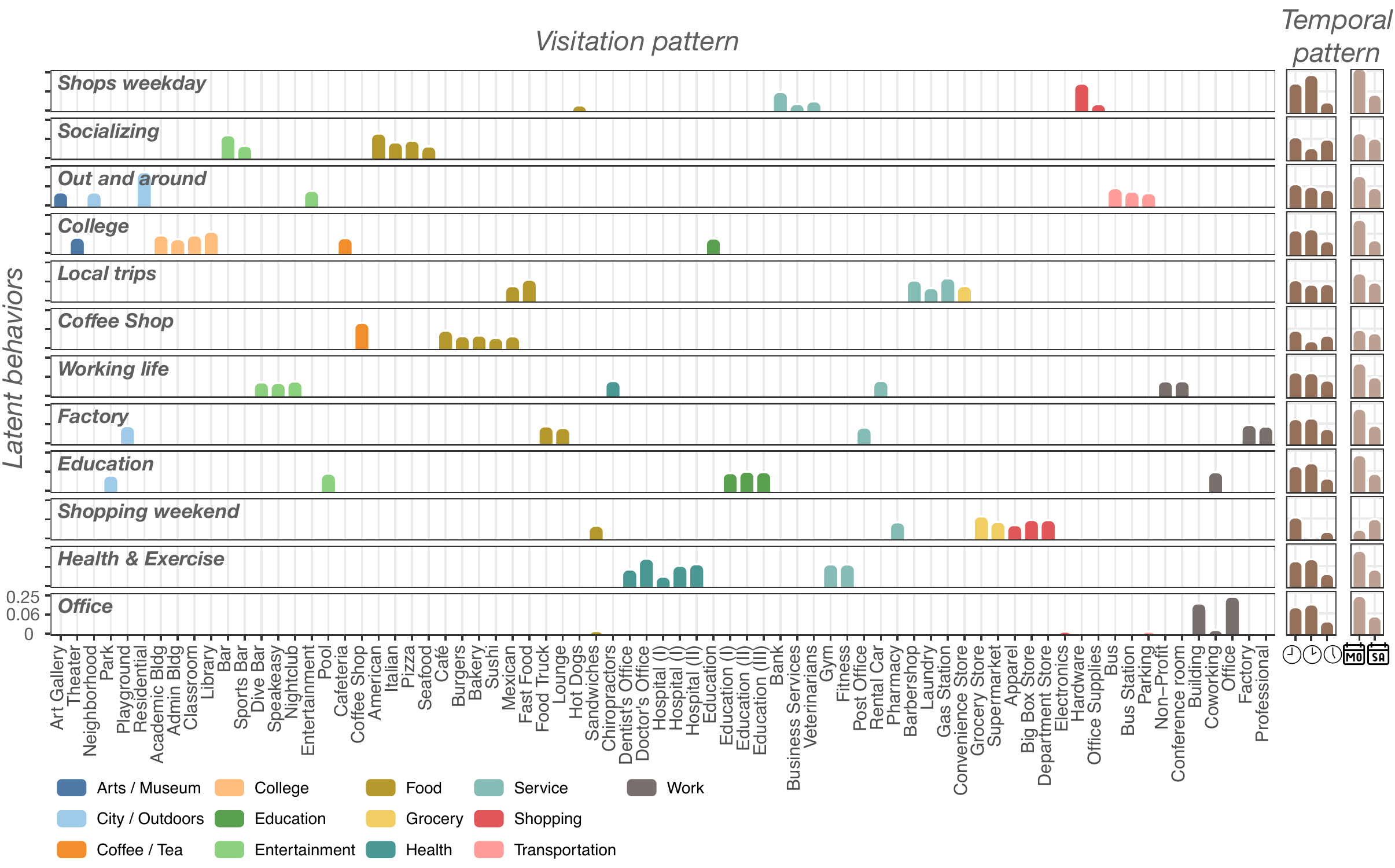}
\caption{{\bf Latent behaviors:} Visitation and temporal components for each of the $k=12$ latent behaviors detected. For simplicity, only the top 7 venue category components by behavior are shown. Colors correspond to the different classifications of the venues. Temporal patterns correspond to the fraction of morning, afternoon, and night visits together with the fraction of weekday and weekend visits. The ``Out and around'' main components are residential and bus transportation during weekdays, while ``Socializing'' is composed of visitations to bars, sports bars, and food places like Italian, American, pizza or seafood, especially during the evenings/nights. Finally ``Local Trips'' refers to a latent behavior mainly composed of running errands (laundry, gas station, convenience store) and visitation to fast food or food truck venues.}
\label{figure2}
\end{figure*}

The number of latent behaviors $k$ is obtained using standard error methods and bi-cross validation for NMF \cite{bicrossOwen,aledavood2022,mollgaard2022} (see Methods and \smref{sectionSM:rankselection}), to get $k = 12$ latent behaviors. The moderately large number of latent behaviors shows the richness and heterogeneity of our dataset. To interpret the latent behavior patterns, we first look into the different categories' dominance values and time slots (see Figure \ref{figure2}). As we can see, most of the latent behaviors are easily recognizable and, as expected \cite{hu2016tales}, their most relevant components belong generally speaking to combinations of working, food, entertainment, or shopping activities. Note that they are not strict projections only on one of those dimensions. For example, we find a latent behavior (``Working life'') of working-related activities (Conference Room, Non-Profit) and nightlife venues, or a latent behavior ``Out and around'' that combines public transportation (bus) with neighborhood visits. Our choice of including the daily and weekly temporal pattern allows us to detect even different shopping behaviors between weekends (``Shopping weekends'' that also includes grocery shopping) and weekdays (``Shops weekdays''). Other distinct latent behaviors correspond to ``College'' students, ``Coffee shop'' frequenters, or ``Health \& Exercise'' visitors. Note that our denomination of the latent behaviors is based on the most dominant categories, which are also the most visited categories in cities. This does not mean that other less-visited categories are not part of those behaviors. For example, most latent behaviors have some components in the Food category (see \sfref{figSM:allcomponents}). However, their relative importance is smaller than in the ``Local trips'', ``Coffee Shop'' or ``Bar + Food'' latent behaviors. Nevertheless, even for the places that are visited less frequently in our cities, the NMF can detect distinct patterns there. For example, the ``Coffee Shop'' latent behavior has large components in airport transportation venues than the rest of the behaviors (see \sfref{figSM:allcomponents}). Finally, it is worth noticing that our detected latent behaviors are not only related to expenditure: an analysis based only on expenditure patterns would have probably missed important latent behaviors like ``Out and around'', ``Office'', ``College'' or ``Education''.

By definition, our latent behaviors encode those visitation patterns which are more frequently happening together. For example, we find in the ``Local trips'' latent behavior that fast food consumption is related to local errands, while nightlife is associated with work-related places like Conference Rooms or Rental Cars. ``Shopping weekend'' tells us that people tend to bundle visits to pharmacies, retail, groceries or departmental stores together. At the same time, in ``Coffee Shop'' we see that visits to coffee shops are associated with the consumption of some types of food like Bakery, Sushi, and Burgers. Thus, our latent behaviors also tell us how people organize their mobility visitation patterns in their daily lives.

\begin{figure*}[t]
\centering
\includegraphics[width=0.95\textwidth]{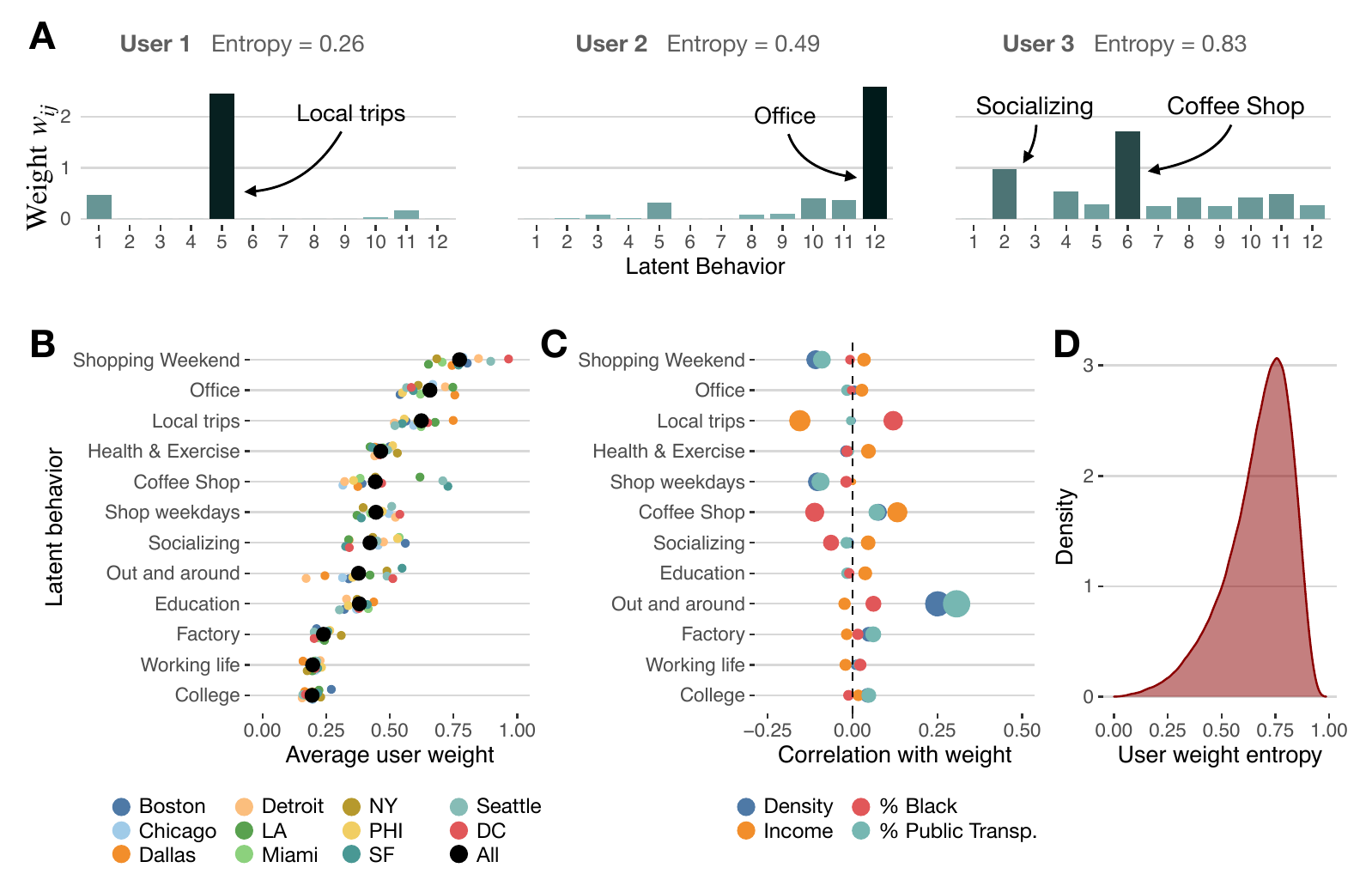}
\caption{{\bf Users' latent behaviors and lifestyles} A) Distribution of weights by latent behavior for three different users in our database. User 1 behavior is dominated by the  ``Local trips'' behavior and has low entropy. However, user 3 has a larger entropy, and her behavior is a very diverse mixture of latent behaviors, including ``Bar + Food'' and ``Coffee Shop''. B) Average weight for the different latent behaviors in all areas (black) and different cities. C) Correlation between the weight of latent behaviors and different demographic and urban characteristic features like population density, median income, percentage of Black population, and percentage of people using public transportation. D) Distribution of user's weight entropy.}
\label{figure3}
\end{figure*}

As we said, each person's activity lifestyle can be described as a linear combination of the latent behavior patterns according to their weights obtained by NMF (see Figure \ref{figure3}). Instead of being described as a simple latent behavior, we find that a user's lifestyle generally depends on many behavior patterns. This is shown by the large entropy of weights by user $S_i  =  -\sum_{j=1}^k \hat w_{ij} \log \hat w_{ij} / \log k$ (where $\hat w_{ij}$ is the normalized weight by user), see Figure \ref{figure3}c. Strict dominance of single latent behavior would make $S_i=0$, while we get $S_i=1$ if all latent behaviors are present and equally important. In our data, we obtained that the average entropy is $\overline{S}_i = 0.67 \pm 0.15$, and thus many latent behaviors configure each user's activity pattern.

However, some latent behaviors are, in general, more frequent than others. For example we find that the average weight of ``Shopping Weekend'' ($\overline{w}_{ij} = 0.774 \pm 0.001$) or ``Office'' ($\overline{w}_{ij} = 0.657\pm 0.001$) is larger than others like ``Working life`` ($\overline{w}_{ij} = 0.197 \pm 0.001$) or ``College'' ($\overline{w}_{ij} = 0.195 \pm 0.001$), which signals that, as expected, the former latent behaviors are more common in urban areas than the latter ones. That is, most individuals have a very small (or even zero) ``College'' latent behavior, while most people have some weight on ``Shopping Weekend'' behavior.

Interestingly, we find that these results are pretty robust across all the cities studied (see Figure \ref{figure3}B), which means that each latent behavior's relative weights are very similar despite the different geography, density, or even cultural nature of the cities. This is an important result that demonstrates our method's robustness across different cities and the homogeneity of activity patterns across the U.S. However, there are small but important variations. For example, the relative weight of the ``College'' behavior is larger in Boston $\overline{w}_{ij}^{\mathrm{Boston}}= 0.270\pm 0.001$ because of the large population of university students in the area. Also, cities with better public transportation systems like Boston, Washington DC, or N.Y., have larger weights in the ``Out and around`` latent behavior than cities like Dallas or Detroit, where public transportation is scarce. Other significant variations happen in the ``Coffee Shop'' latent behavior, more present in cities like San Francisco or Seattle than in the rest. This is expected given that those cities are the ones with the most coffee shops per capita \cite{coffeeguide}. Taking together these results shows the adaptability and robustness of our latent behaviors to describe the similarities and peculiarities of activity patterns in the different cities in the U.S.

\subsection*{Latent behaviors are not fully described by demographics or urban characteristics}
One important question is whether the detected latent behaviors can be explained by individuals' simple demographic or urban characteristics. Group segmentation using census-type only data is traditional in marketing, \cite{Spielman:2015ec,esri2018} and even the U.S. Census Bureau used it to design its 2020 campaign \cite{census}. Since mobility is highly influenced by socioeconomic status, access to public transportation, or the density of the urban area, we could expect that the detected latent behaviors might depend strongly on those features. However, we find that users' weights $w_{ij}$ are largely independent of the median income, population density, the fraction of Black population, or even the fraction of people using public transportation (see Figure \ref{figure3}C and \stref{tableSM:weights})). We only find moderate correlations with some latent behaviors. For example, low-income people have more ``Local trips'' latent behavior, while ``Coffee Shop'' is a behavior more likely to be found in high-income areas. Of course, ``Out and around'' latent behavior is more prevalent in areas with higher public transportation use. Apart from those cases, the correlation between our latent behaviors' weights and demographic and urban features is very small $R^2 \leq 0.1$ (see \smref{sectionSM:models} and \stref{tableSM:weights}). Thus, latent behaviors detected using mobility data are different from the traditional census demographic traits. The detected activity patterns give a different and complementary perspective of our cities than traditional census analysis, allowing us to construct a richer {\em behavioral census} that includes those behaviors.

\subsection*{Association of latent behaviors with social, mobility, and health problems}
To demonstrate the complementary power of the latent behaviors to traditional census approaches, we have analyzed their association with different social, mobility, and health outcomes in the 11 cities. In the social dimension, we have considered $I_i$, the income integration (or diversity) experienced by each individual, introduced in \cite{moro}. This quantity reflects how homogeneous is the exposure of each individual to the different income groups in the city: by using the household median income of the Census Block group where user $i$ lives, we can quantify the income group (income quartile within each city) she belongs to. Using that information for each user, we can estimate the amount of time a user $i$ is exposed to the different income groups in the city while visiting different venues: if $I_i = 0$, individual $i$ only goes to places where her particular income group is the majority. If $I_i = 1$, the individual is exposed equally people from all the city's income groups (see Materials and Methods). Other versions of diversity exposure have been analyzed recently \cite{athey} and, in particular, income exposure diversity is related to social capital and impacts economic opportunities and social income mobility of individuals \cite{chetty}. Also, along the social dimension, we have studied the individual's place exploration $E_i$, which measures the rate of visitation to different places by $i$ in our time period \cite{Alessandretti:2018jn}. Although people spend most of their time in a very small number of places \cite{Song:2012wk,Gonzalez:2009p1717}, it is well known that some tend to visit more places ({\em explorers}, $E_i \simeq 1$), while some others spend most of their time in a small set of places ({\em returners}, $E_i \simeq 0$). These social metrics are crucial to understanding the social component of mobility in our cities and, specifically, how segregated (not integrated) are people living in them. As was found in \cite{moro} experienced income integration is moderately and positively related to place exploration ($\rho = 0.456 \pm 0.001$). To test the association of the latent behaviors in these problems, we have used a regression model:
\begin{equation}
\label{modeleq}
    I_i, E_i \sim \sum_{j=1}^k \beta_j w_{ij} + \sum_{l=1}^m \gamma_l d_l + \mathrm{MSA}_i + \varepsilon_i
\end{equation}
where $d_l$ refers to the four demographic and urban features mentioned before (median household income, the density of the area, the fraction of Black people, and the fraction of use of public transportation), and $\mathrm{MSA}_i$ is a fixed factor by city (Metropolitan Statistical Area). Including the census variables and city-fixed effects allows us to investigate the fundamental role of latent behaviors once we control for potential effects by demographic and urban characteristics and the city where users live. In our models, census features are always less important to explain that variability ($R^2 = 0.059$ for $I_i$ and $R^2 = 0.025$ for $E_i$ only using census variables) than our latent behaviors ($R^2 = 0.164$ and $R^2 = 0.26$ respectively using also latent behaviors, see \stref{tableSM:outcomes} and \stref{tableSM:outcomesdemo}). This result shows that our latent behaviors encode most of the social-economic integration and exploration variability across users, which are largely independent of census variables.

\begin{figure*}[t]
\centering
\includegraphics[width=0.95\textwidth]{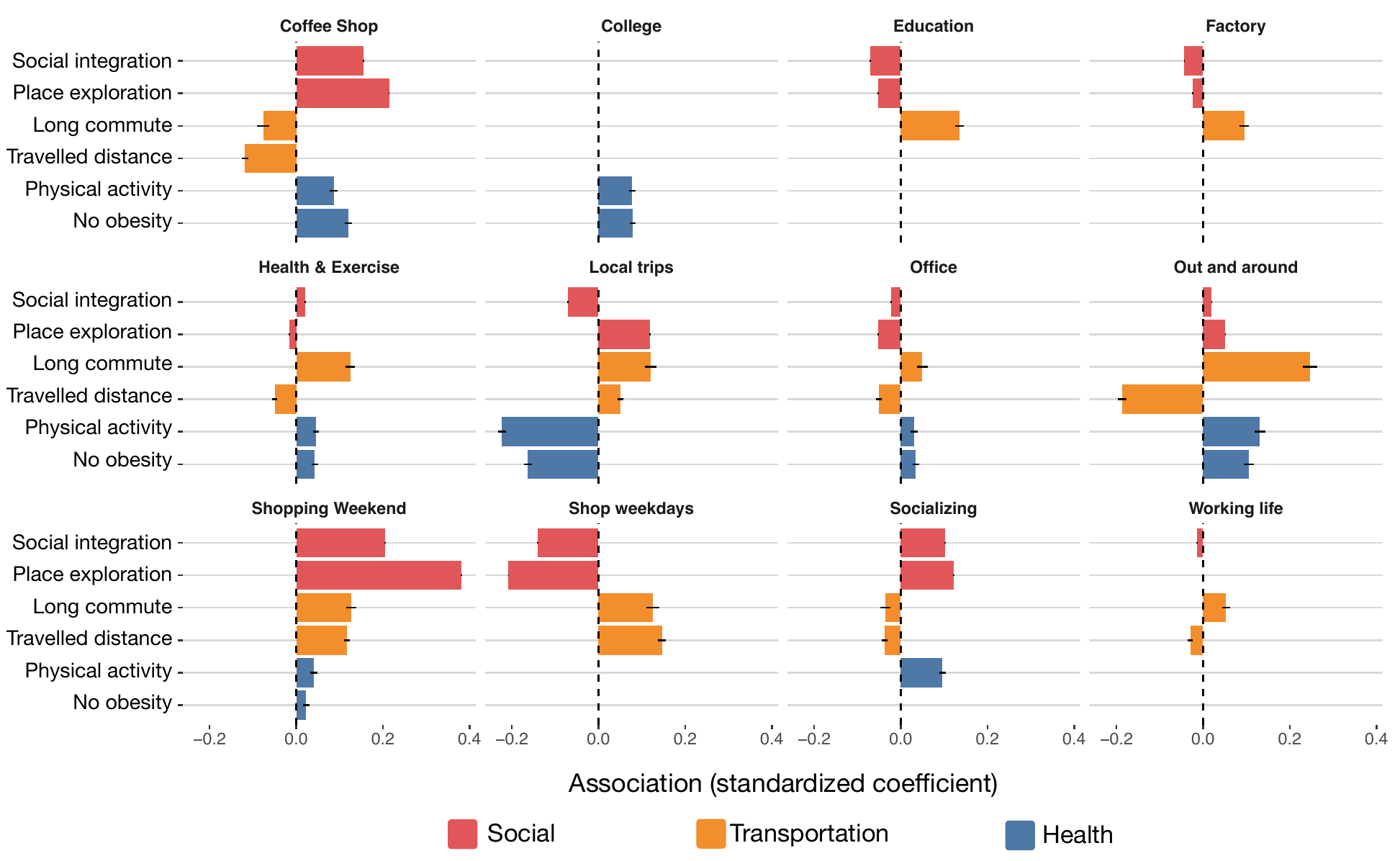}
\caption{Relationship of behavioral patterns with social, mobility, and health problems. Bars show the coefficient of each behavior weight on the different regression models for social-economic integration, place exploration, the fraction of people with longer than 45 min commute, amount of distance traveled by day, the fraction of people doing daily physical activity, and fraction of people not having obesity. }
\label{figure4}
\end{figure*}

Nevertheless, not all latent behaviors have the same effect on experienced income segregation and exploration. Figure \ref{figure4} shows the relationship [measured as the standardized coefficient $\beta_j$ in model (\ref{modeleq})] of the different latent behaviors with both social problems. We find that the latent behaviors that impact economic integration are related to shopping or food/coffee. In contrast, others like college, office, or health are not heavily associated with economic integration. Interestingly, behaviors like ``Shopping Weekend'' or ``Coffee Shop'' are positively associated with economic integration, while behaviors ``Education'', ``Factory'' or ``Shopping weekdays'' are more present in users with more considerable experienced income segregation. This might be explained by the fact that shops, coffee shops, or some restaurants are more economically diverse in the city than factories, education, or local shops \cite{moro}. In general social exploration follows the same pattern, although people with more ``Local trips'' behavior tend to be more explorers without being more integrated. These results show that the latent behaviors carry significant explanatory power of the income diversity and exploration experienced by people in the urban areas analyzed.

We have also studied other problems related to transportation and health. Lifestyles are crucial to understanding our mobility choices or opportunities and transportation routines \cite{Salomon.1983}, but also physical activity and the prevalence of some health conditions \cite{joumard2010health}. Since we do not have individual health conditions, we have used data from the Census, the Bureau of Transportation Statistics (BTS), and the Center for Disease Control to take variables by census tract $\alpha$ describing the average distance traveled by residents $D_\alpha$, the fraction of people that have more than 45 minutes of commuting $C_\alpha$, the fraction of people with leisure-time physical activity in the past month $P_\alpha$ and the fraction of people with no obesity $O_\alpha$ (see Materials and Methods). To study the relationship of our latent behaviors with those problems, we construct the average weight by census tract $\hat w_{\alpha,j}$ for all the users living in that tract $\alpha$ and fit them using similar models as Eq. \ref{modeleq} (see \smref{sectionSM:models}).

As we can see in Figure \ref{figure4}, local behaviors like ``Out and around`` have a strong negative association with the distance traveled, although the association with commuting duration is positive. In general, tracts with more shopping behavior travel more, while those with larger ``Coffee Shop'' or ``Bar + Food'' latent behavior have smaller commutes and distance traveled. Since our data is projected at the level of the census tract, individual variability is averaged out, and thus, demographic and urban variables are more important to explain the variability in transportation variables. However, our latent behaviors still explain part of how people commute or move around the city, even if we condition on income, race composition, density, or use of public transportation (see \smref{sectionSM:models}).

Finally, we see in Figure \ref{figure4} that, although not all of them are relevant, there are a fraction of latent behaviors that have a significant association with health outcomes. As expected, more presence of latent behaviors like ``Local trips'' (which include visits to Fast Food venues) is associated with less Physical Activity and more Obesity, while behaviors like ``Out and around'' or ``Coffee Shop'' are positively associated with the amount of Physical Activity and the absence of Obesity. However, socially and mobility important behaviors like shopping are not significantly related to health outcomes. This is even after controlling for demographic variables like income or race, which are the most critical determinants of those health outcomes \cite{obesitycdc}.

In summary, our results show that our latent behaviors that constitute individual lifestyles are significantly correlated with social, transportation, and health outcomes. Different behaviors are related to different dimensions, and in some cases, the importance of latent behaviors is similar to or even surpasses that of demographic variables. For example, knowing that a particular user has extensive shopping or food/coffee behaviors can better explain her economic integration and exploration than knowing her income (see \stref{tableSM:outcomes} and \stref{tableSM:outcomesdemo}). Or conditioning on income or population density, we see that some behaviors like ``Out and around'' or ``Coffee Shop'' are associated with transportation and health outcomes.


\section*{Discussion}
Understanding urban problems require a good description of human behavior in cities \cite{Batty:2013cv,Cagney.2020}. Our research shows that the high-dimensional nature of mobility of millions of people visiting millions of places in the U.S. can be projected onto a small set of latent behaviors that capture their routines and habits that result from their choices or opportunities accessible to them. Demographics or urban characteristics cannot fully explain those latent behaviors, and people living in the same neighborhood with the same income, race, or educational levels might have different shopping, working, or leisure latent behaviors, resulting in entirely different lifestyles. Since the composition of the lifestyles is robust across different geographical areas and cities, our results could be used to build characterizations and compare individuals and groups at different geographical and demographic levels. This could enrich the current Census by including the composition of the different latent behaviors to study urban areas. It could also help in methods of exploiting mobility data by preserving the privacy of individuals. This can be done by computing projected aggregated variables along those latent behaviors rather than detailed and more invasive individual visitation patterns.

Our latent behaviors describe how people organize their visitation patterns and mobility around the city. For example, we find that people that make trips to errands also visit fast food outlets frequently (``Local trips'' latent behavior), while heavy users of public transportation (bus) also spend much time in the neighborhood and entertainment (``Out and around''). Working life is also related to nightlife. These dependencies show that those latent behaviors represent combined aspects of our life that occur concurrently and which could be used to devise successful holistic interventions to change people's lifestyles. For example, people that run many errands might choose fast food because they are time-poor or because errands take place around specific food environments (food swamps). Our results can help design public health interventions that incorporate those distinct lifestyles to identify those routines and habits that are most risky for health \cite{habits2011}.

We note that latent behaviors have a different relationship with social, transportation, and health outcomes. For example, while weekend shopping behaviors are associated with a more exposure to economic diversity of urban dwellers, they carry more commuting time and travel distance and thus more pollution. Similarly, the ``Out and around'' latent behavior is associated with longer commutes and more physical activity or the absence of obesity. Since most urban interventions are likely to change the relative weight of those latent behaviors or ultimately change them completely, it is essential to balance the trade-off among social, transportation, and health outcomes encoded in those behaviors. Also, not all behaviors have the same weight in describing users' lifestyles. Shopping, food, or working latent behaviors are the most important, suggesting that they are the ones where interventions to change experienced income segregation, transportation, or health outcomes could be more considerable \cite{zhao2020}. 

Our results show that activity lifestyles are not monolithic groups of homogeneous behavior among people. Our framework of describing lifestyles as a combination of latent behaviors reflects that lifestyles are instead a continuum spectrum of the relative balance between work, shopping, transportation, or leisure time. Given the ubiquitous nature of mobility and activity data from mobile phones, we hope this framework could be used in the future to understand better the rapid and extensive scale changes in other urban areas and cities worldwide.

\section*{Methods}

\subsection*{Mobility Data}
To obtain the number of visits to different categories of places, we first extract the places where users stay for more than $5$ minutes based on the user's location and timestamp. Then, all the places are classified into different categories according to the Foursquare Category Hierarchy \cite{catsfsq}. For each user, we construct a vector $\mathbf{x}_i$, which contains the number of visits by category, together with the fraction of total visits during the morning, afternoon, evening/nighttime, weekdays, and weekends. The number of visits by category is normalized for each user. We only consider users that have more than 50 visits in the six-month period. Only users with more than 50 visits during the period and with at least 5 categories visited were considered. Finally, we only considered the top most visited 238 categories to prevent over-fitting to small, infrequent categories. 

Due to the anonymous nature of our location data, we obtained the demographic characteristics of each user at the area level. To this end, we used the most visited Census Block Group \cite{acs} during the night to estimate the home location of each user. See \smref{sectionSM:data} for more details about the data.

\subsection*{Other data}
Demographic data like median household income, the fraction of Black population, the fraction of people that use public transportation, or urban characteristics like population density were obtained from the Census 2013-2017 ACS 5-year Estimates \cite{acs}. The fraction of people with more than 45 minutes of commuting was also obtained from the ACS Data. The source of transportation data is the 2017 Local Area Transportation Characteristics for Households Data done by the Bureau of Transportation Statistics (BTS) \cite{bts}, while obesity prevalence and physical activity are given by the 500 Cities Project Data from the Center for Disease Control (CDC) \cite{cdc}. Obesity prevalence is measured as the percentage of adults, aged 18 or older, who report a body mass index (BMI) of 30 or higher. Physical activity is measured as the fraction of adults who report getting leisure-time physical activity in the past month. Cities are defined as the Census Core Based Statistical Areas \cite{cbsa} that are socioeconomically metropolitan areas related to an urban center. Note that although we could have constructed the mobility variables using our data, we used the BTS and Census data because the latter is based on more reliable estimation statistics. But also because our data do not have complete daily individual trajectories of people, preventing us from having a precise estimation of the distance traveled and the commuting time.

To measure experienced social-economic integration, we use the inequality metric introduced in \cite{moro} to estimate how unequal is the exposure of an individual to the different income groups in the city. To this end, we divide the sample of users in each city into four quartiles according to the median household income of their home Census Block group \cite{acs}. Social-economic integration was measured as $I_i = 1-  \frac23\sum_q|\tau_{iq}-1/4|$, where $\tau_{iq}$ is the proportion of time user $i$ is exposed to group $q$ of income. That proportion is calculated by looking at the weighted distribution of income of the people that $i$ encounters in the places she visits. Specifically $\tau_{iq} = \sum_\alpha \tau_{i\alpha}\tau_{q\alpha}$, where $\tau_{i\alpha}$ is the fraction of time that $i$ spends at place $\alpha$ and $\tau_{q\alpha}$ represents the proportion of time at place $\alpha$ spent by income group $q$. Our metric for individual economic integration can be thought of as an extension of the traditional metric of isolation or interaction for groups to the level of individuals based on daily encounters among them.  Finally, social exploration is measured as $E_i = S_i / N_i$, where $S_i$ is the total number of different places visited by $i$ and $N_i$ is the total number of visits to places by $i$. See \cite{moro} for more details on these metrics and their distribution.

\subsection*{Non-negative matrix factorization}
The activity matrix $X$ is factorized as $X = W \cdot B$ using different ranks $k$. To do that, we used non-negative matrix factorization (NMF) using fast sequential coordinate-wise descent and Kullback-Leibler divergence for the loss function. We run the (NMF) one hundred times for each value of $k$. Different methods were used to assess the value of $k$ (see \smref{sectionSM:rankselection}), including bi-cross validation \cite{bicrossOwen}. We found that $k = 12$ was the one that optimizes the error, and the stability of the weights $W$ while making the latent behaviors $B$ more interpretable across realizations. For that $k=12$, we chose $B$ and $W$ from the realization with smaller KL loss (see \smref{sectionSM:rankselection}).

To prevent an over-representation of the larger cities in the factorization, we did not use our 1.2 million users in the NMF. Instead, we randomly selected 10k users in each city and constructed the matrix $X_0$. We factorize it into $W_0 \cdot B_0$ and use $B_0$ to solve the non-negative linear regression problem $X \sim W \cdot B_0$ to get $W$ for the rest of the users. This way, we get a fair representation of all the latent behaviors commonly present across cities. For comparison, we have also used LDA to detect the latent behavior patterns (see \smref{sectionSM:LDA}). Although the results are somehow similar, we find that the latent behavior patterns detected by NMF have better interpretations than that of LDA. 

\bibliography{activity}

\section*{Acknowledgements}

We would like to thank Cuebiq for granting us access to mobility data through their Data for Good program. We would like to thank MIT ILP program and Ferrovial for their financial support for this project. E.M. acknowledges partial support by MINECO (FIS2016-78904-C3-3-P and PID2019-106811GB-C32).

\section*{Author contributions statement}

All authors designed research. Y.Y. and E.M performed research and analyzed the data. All authors wrote the paper.

\subsection*{Data availability}
The data that support the findings of this study is available from Cuebiq through their Data for Good program, but restrictions apply to the availability of these data, which were used under licenses for the current study, and so are not publicly available. Aggregated data used in the models are, however available from the authors upon reasonable request and permission of Cuebiq.

\section*{Competing interests}

The authors declare no conflict of interest.

\newpage
\appendix

\noindent
\section*{Supplementary Material}
\renewcommand{\figurename}{Supplementary Figure}
\renewcommand{\tablename}{Supplementary Table}

\section{Data}
\label{sectionSM:data}
The mobility data was obtained from Cuebiq, a location intelligence and measurement company. The dataset consists of anonymized records of GPS locations from users that opted-in to share the data anonymously in the Boston metropolitan area over a period of 6 months, from October 2016 to March 2017. Data was shared in 2017 under a strict contract with Cuebiq through their Data for Good program where they provide access to de-identified and privacy-enhanced mobility data for academic research and humanitarian initiatives only. All researchers were contractually obligated to not share data further or to attempt to de-identify data. Mobility data is derived from users who opted in to share their data anonymously through a General Data Protection Regulation (GDPR) and California Consumer Privacy Act (CCPA) compliant framework.

From the data we extracted the ``stays'', as the places where anonymous users stayed (stopped) for at least 5 minutes using the algorithm proposed by Hariharan and Toyama \cite{hariharan2004project}. Some of the stays happen within places (Points of Interest). We use a dataset of 1.2 Million Points of Interest in US metropolitan areas collected using the Foursquare API. We use the Foursquare venue categorization of the places to detect the type of place visited \cite{catsfsq}. Finally we estimate the home Census Block Group of the anonymous users as that in which they are more likely located during nighttime. This results in a dataset of the places people stayed including the points of interest that anonymous users visited and the most likely census block group of where the device owner lives.

We only considered mobility data which happen within 11 metropolitan areas defined as the Core-based Statistical Areas (CBSA) \cite{cbsa}. We considered CBSAs instead of other geographical units, since they are areas that are socioeconomically related to an urban center. This provides a self-contained metropolitan area in which people move for work, leisure or other activities. Note that most of the CBSAs we consider span several states.

\section{Non-negative matrix factorization}
Non-negative matrix factorization is a well-known technique to approximate a non-negative matrix $X\in[0,\infty]^{m\times n}$ using non-negative low-rank matrices $W \in [0,\infty]^{m\times k}$ and $H \in [0,\infty]^{k\times n}$ such that $X \simeq W H$. This is typically achieved by looking for the $W$ and $H$ that minimize a loss function. Typical choices are the quadratic loss $||X - W H ||_F^2$, where $||\cdots||_F$ is the Frobenius norm, or the Kullback-Leibler divergence distance $\mathrm{KL}(X | WH)$, where
\begin{equation}
    \mathrm{KL}(A,\hat A) = \sum_{ij} a_{ij}\log \frac{a_{ij}}{\hat a_{ij}} - a_{ij} + \hat a_{ij}.
\end{equation}

Different algorithms exists to approximate the minimum of the loss function get $W$ and $H$. In our case we have used the fast sequential coordinate-wise descent introduced by Lin and Boutros \cite{lee} and implemented in the {\tt NNLM} package in {\tt R} \cite{nnlm}. For the loss function we used the Kullback-Leibler divergence metric.

Note that since the non-negative matrix factorization is not guaranteed to be unique, each factorization slightly depend on the different initial condition. Thus, for each rank $k$ we have run 200 realizations. For presentation purposes, the realization with smallest error by $k$ was used in the definitions of the latent behaviors weights $w_{ij}$ used here and in the main paper.

\section{Rank selection}
\label{sectionSM:rankselection}
There are several strategies to choose the rank $k$ in the non-negative matrix factorization (NMF). As in other latent detection methods, typically a combination of statistical and interpretability criteria is used \cite{GraellsGarrido:tq,mollgaard2022,aledavood2022}. Popular statistical methods include observing the variation of the residual sum of squares (RSS) between the original matrix and its factorization \cite{hutchins2008} or measuring the stability of the weights in the NMF \cite{brunet2004} across different realizations. Other state-of-the-art methods include bi-cross validation methods \cite{bicrossOwen}, where a set of rows and columns are left out to evaluate the goodness of the factorization to reconstruct them. But we also would like to get latent behaviors which are interpretable, that is, that can be described by a number of significant components (not too sparse) and have some meaning according to human interpretation. 

In our case, we have chosen a combination of the two approaches. To measure the sparseness of each latent behaviors $\mathbf{b}$ we have used the entropy
\begin{equation}
S(\mathbf{b_j}) = -\frac{1}{M}\sum_{l=1}^M \frac{b_{jl}}{||\mathbf{b}_j||_1} \log \frac{b_{jl}}{||\mathbf{b}_j||_1}. 
\end{equation}
Entropy is zero if $\mathbf{b}_j$ only contains a non-zero component and one if all components are equal. For each $k$ we have computed the average of all behaviors and all realizations. Although sparsity of $W$ and $B$ can be fixed by imposing some additional constrains in the factorization process \cite{hoyer2004}, we have preferred not to used them to simplify the process of rank selection by minimizing the number of choices made and parameters used in the factorization. 

For the bi-cross validation, we have used the method by Owen {\em et al.} \cite{bicrossOwen}. In that method, a set of randomly-selected $r$ rows and $s$ columns are removed from the original matrix $X$. Rearranging the matrix in the following form
\begin{equation}
    X = \left(\begin{array}{cc} A & B \\ C & D \end{array} \right)
\end{equation}
where $A$ is the $r \times s$ submatrix. Assuming that $D = W_D H_D$ is the rank $k$ non-negative factorization of $D$, we can use approximated the substracted matrix $A$ by 
\begin{equation}
    \hat A = U_B V_C
\end{equation}
where $U_B$ is the solution of the non-negative least squares problem $U_B = \arg \min_W \mathrm{KL}(B | W H_D)^2$ and $V_C = \arg \min_H \mathrm{KL}(C | W_D H)^2$. Then the held-out error estimate for the cross-validation is $\mathrm{KL}(A|\hat A)$. We have repeated this bi-cross validation process for $300$ random selections of $r$ rows and $s$ columns. Specifically we have left 10\% of the rows and columns in each random selection.

Finally, to measure the stability of the components, we have used a variation of the cophenetic correlation coefficient proposed by Brunet et al. \cite{brunet2004}. For each realization we assign each user $i$ to the cluster given by her largest component $w_{ij}$ and compare those assignments across different realizations using the normalized mutual information (NMI) measure proposed by Danon et al \cite{Danon:2005p412}. Finally we average that NMI over all possible comparisons between realizations. If all realizations give the same clusters then our metric will be one. However, if clusters are totally different across realizations, the NMI will be zero.

\begin{figure}
\centering
\includegraphics[width=\textwidth]{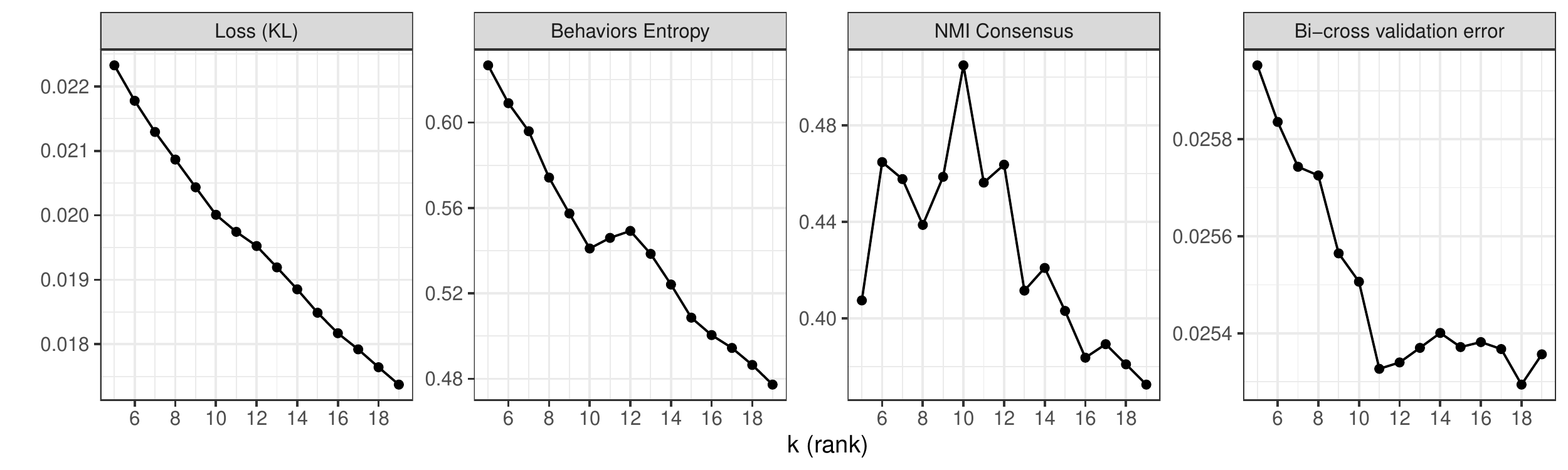}
\caption{{\bf Rank selection metrics.} Plots show the average values of the entropy for all behaviors, the bi-cross validation error, the average loss (KL divergence) and average consensus (NMI) across 100 realizations for each factorization rank $k$.}
\label{figSM:rankselection}
\end{figure}

Figure \ref{figSM:rankselection} shows the variation of the (average) loss (KL divergence) and entropy of the latent behaviors as a function of NMF factorization rank $k$. Although the KL loss curve does show only a small inflection point around $k=12$, we can clearly see that the sparseness of the latent behaviors has a local minimum around $k=12$. We also find that the consensus NMI metric has a maximum at $k=10$ and local one at $k=12$. Finally, the bi-cross validation seem to have a local minimum around $k= 11, 12$. Given these results and the good interpretability of the latent behaviors, we chose $k=12$ as the factorization rank in our NMF. Similar protocol to choose the rank have been used in similar datasets recently \cite{mollgaard2022,aledavood2022}.

\section{Comparison with LDA}
\label{sectionSM:LDA}
We have studied also the detection of latent behaviors using other methods like Latent Dirichlet Allocation (LDA). As we can see in Figure \ref{figSM:LDA}, the topics detected using LDA on the matrix $X_0$ are similar to the ones obtained using NMF. However, some of the behaviors in the NMF get mixed in the LDA, specially the ones around shopping, school or health, which make them less interpretable. Apart from these differences, this result shows that our detected latent behaviors are genuinely present in the data and that different methods yield more or less to the same set of them.

\section{Models}
\label{sectionSM:models}
To detect the dependence between our latent behaviors, demographic variables, and social, transportation, and health outcomes we have used Ordinary Least Squares (OLS) regressions between the variables at individual or census tract level. In particular:
\begin{itemize}
    \item To test the independence between the latent behaviors and the demographic variables we have model $w_{ij}$, each individual $i$ weight on latent behavior $j$ as
    \begin{equation}
    \label{eqSM:weights}
    w_{ij} \sim \mathrm{density}_i + \mathrm{income}_i + \mathrm{public\_transportation}_i+\mathrm{black\_population}_i + \varepsilon_i
    \end{equation}
    where each demographic variable is the estimation from the 2013-2017 American Community Survey 5-year Estimates \cite{acs} for the Census Block Group where we estimate where the use lives. Results for this OLS regression are presented in Table \ref{tableSM:weights}, where we can see that the $R^2$ is clearly very small.
    \item To test the impact of latent behaviors on social outcomes we have used the OLS regression at individual level:
    \begin{equation}
    \label{eqSM:model1}
    X_i \sim \sum_{j=1}^k \beta_j w_{ij} + \mathrm{density}_i + \mathrm{income}_i + \mathrm{public\_transportation}_i+\mathrm{black\_population}_i + \varepsilon_i
    \end{equation}
    where $X_i$ is each individual's social integration or exploration. Results are presented in Table \ref{tableSM:outcomes}.
    \item Finally, since we can only access transportation and health outcomes at the Census Tract level, we test the impact of latent behaviors on those outcomes using OLS at census tract $\alpha$.
    \begin{equation}
    \label{eqSM:model2}
    X_\alpha \sim \sum_{j=1}^k \beta_j \hat w_{\alpha j} + \mathrm{density}_\alpha + \mathrm{income}_\alpha + \mathrm{public\_transportation}_\alpha+\mathrm{black\_population}_\alpha + \varepsilon_\alpha
    \end{equation}
    where $X_\alpha$ is each of the variables studied (no obesity, physical activity, long commutes or distance travelled) by census tract $\alpha$ and $\hat w_{\alpha j}$ is the average of $w_{ij}$ for all the users living in census tract $\alpha$. Results are presented in Table \ref{tableSM:outcomes}
\end{itemize}
Finally, we have build also models for $X_i$ and $X_\alpha$ using only demographic variables to test the relative importance of the latent behavior weights. Results are presented in Table \ref{tableSM:outcomesdemo}

\begin{figure}
\centering
\includegraphics[width=\textwidth]{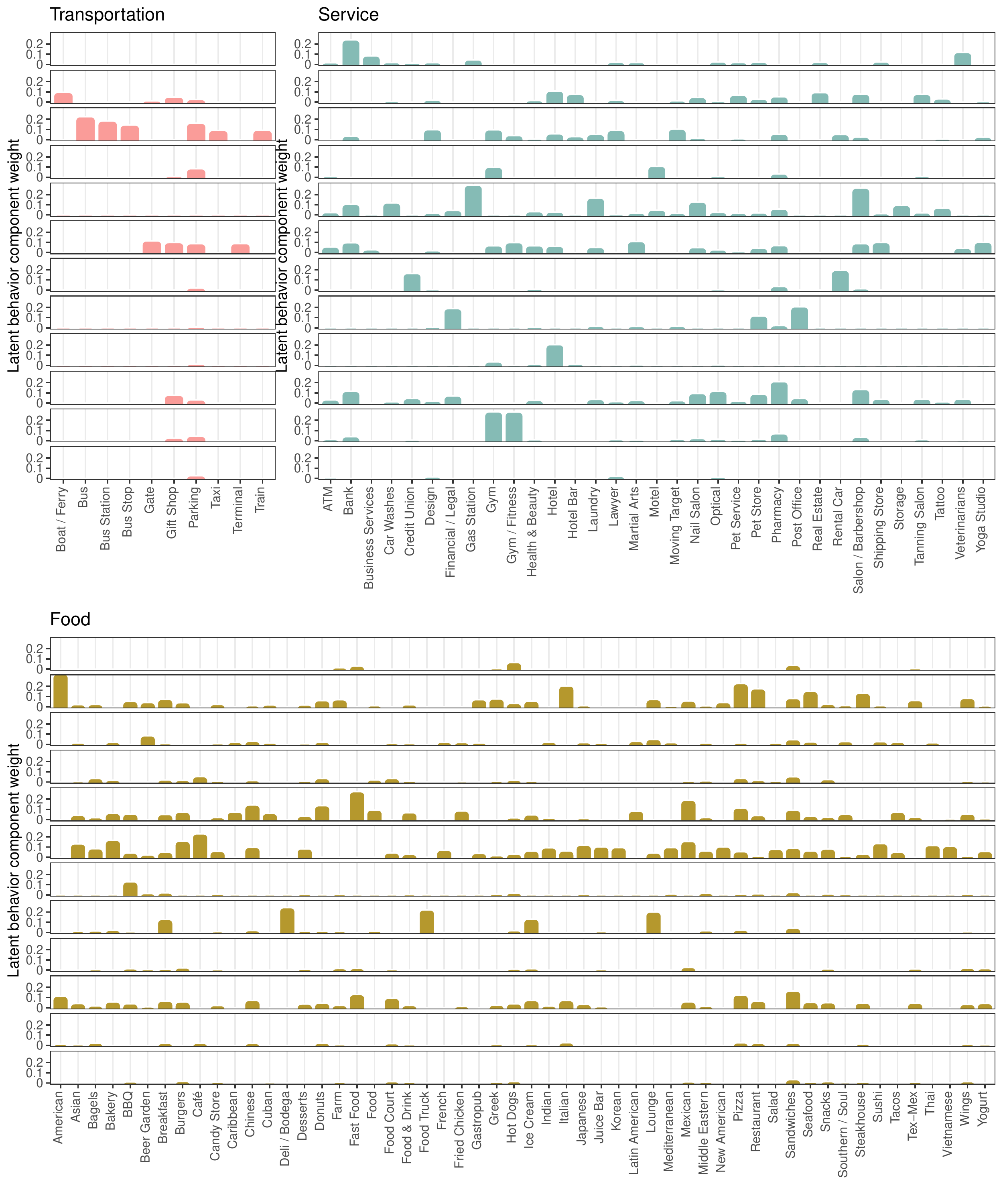}
\caption{{\bf Latent Behavior components:} Panels show all the components for the $k=12$ latent behaviors in the different categories for 3 different type of places: Transportation, Service and Food.}
\label{figSM:allcomponents}
\end{figure}

\begin{figure}
\centering
\includegraphics[width=\textwidth]{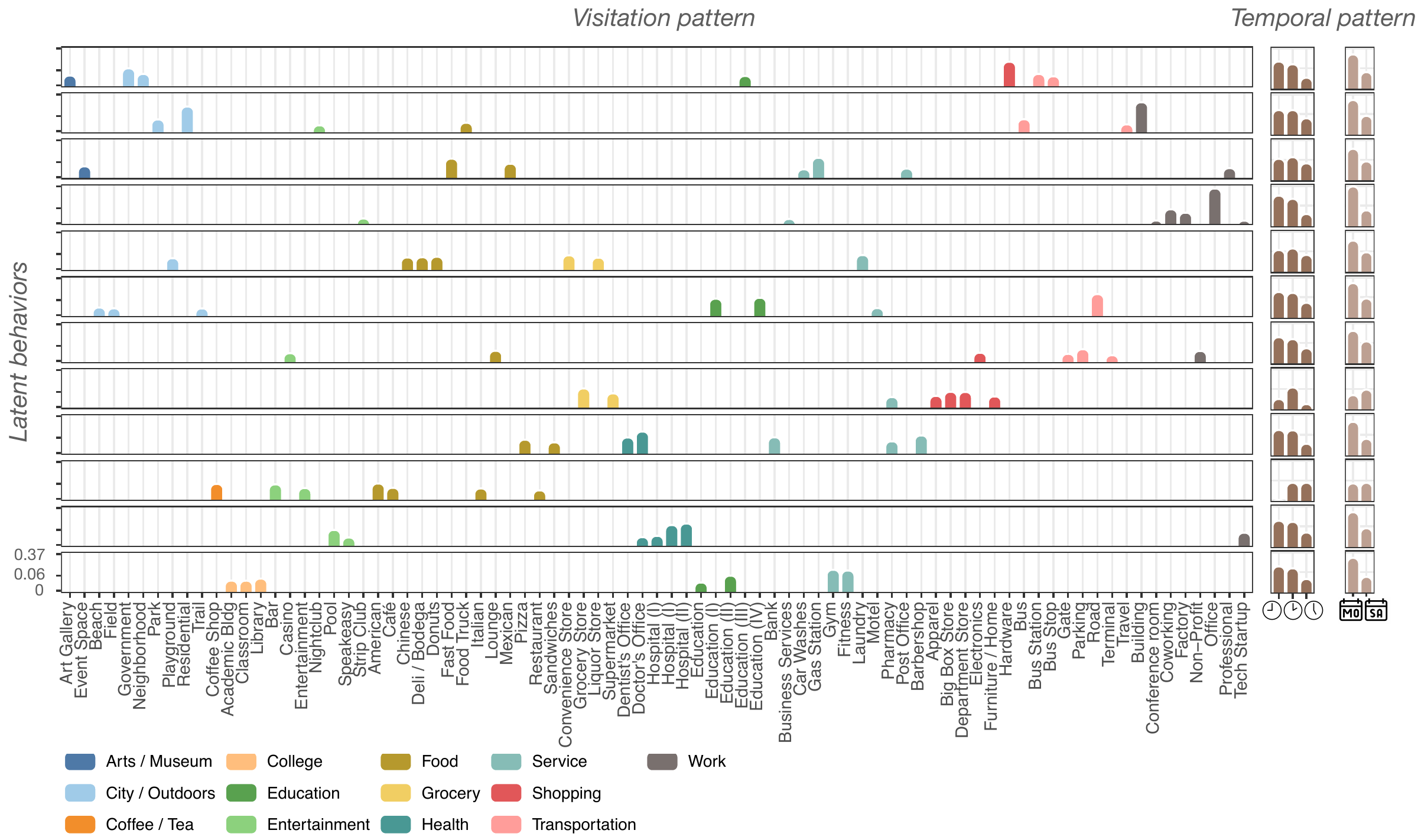}
\caption{{\bf Latent Behavior components using LDA:} Activity and temporal components for each of the $k=12$ latent behaviors detected using LDA. For simplicity, only the top 7 categories by latent behavior are shown. Colors correspond to the different classification of the venues. Temporal patterns correspond to fraction of morning, afternoon and night visits together with fraction of weekday and weekend visits.}
\label{figSM:LDA}
\end{figure}

\begin{table} \centering \small
  \caption{Regression table for latent behavior weights as a function of different demographic variables} 
  \label{tableSM:weights} 
\scalebox{0.73}{
\begin{tabular}{@{\extracolsep{-10pt}}lcccccccccccc} 
\\[-1.8ex]\hline 
\hline \\[-1.8ex] 
 & \multicolumn{12}{c}{\textit{Latent behavior weight $w_{ij}$:}} \\ 
\cline{2-13} 
\\[-1.8ex] & 1 & 2 & 3 & 4 & 5 & 6 & 7 & 8 & 9 & 10 & 11 & 12\\ 
\hline \\[-1.8ex] 
 Density & $-$0.077$^{***}$ & $-$0.002 & 0.105$^{***}$ & 0.026$^{***}$ & $-$0.011$^{***}$ & 0.060$^{***}$ & $-$0.006$^{***}$ & 0.015$^{***}$ & 0.003$^{**}$ & $-$0.078$^{***}$ & $-$0.012$^{***}$ & 0.031$^{***}$ \\ 
  & (0.001) & (0.001) & (0.001) & (0.001) & (0.001) & (0.001) & (0.001) & (0.001) & (0.001) & (0.001) & (0.001) & (0.001) \\ 
  & & & & & & & & & & & & \\ 
 Median household & $-$0.015$^{***}$ & 0.029$^{***}$ & 0.006$^{***}$ & 0.018$^{***}$ & $-$0.132$^{***}$ & 0.115$^{***}$ & $-$0.016$^{***}$ & $-$0.012$^{***}$ & 0.039$^{***}$ & 0.025$^{***}$ & 0.045$^{***}$ & 0.034$^{***}$ \\ 
  income & (0.001) & (0.001) & (0.001) & (0.001) & (0.001) & (0.001) & (0.001) & (0.001) & (0.001) & (0.001) & (0.001) & (0.001) \\ 
  & & & & & & & & & & & & \\ 
 Fraction of users of & $-$0.048$^{***}$ & $-$0.002 & 0.245$^{***}$ & 0.038$^{***}$ & $-$0.029$^{***}$ & 0.067$^{***}$ & 0.020$^{***}$ & 0.052$^{***}$ & $-$0.017$^{***}$ & $-$0.046$^{***}$ & $-$0.0004 & $-$0.036$^{***}$ \\ 
  public transp.& (0.001) & (0.001) & (0.001) & (0.001) & (0.001) & (0.001) & (0.001) & (0.001) & (0.001) & (0.001) & (0.001) & (0.001) \\ 
  & & & & & & & & & & & & \\ 
 Fraction of black & $-$0.006$^{***}$ & $-$0.054$^{***}$ & $-$0.006$^{***}$ & $-$0.018$^{***}$ & 0.087$^{***}$ & $-$0.097$^{***}$ & 0.014$^{***}$ & $-$0.002$^{*}$ & 0.007$^{***}$ & 0.018$^{***}$ & $-$0.001 & 0.014$^{***}$ \\ 
  population & (0.001) & (0.001) & (0.001) & (0.001) & (0.001) & (0.001) & (0.001) & (0.001) & (0.001) & (0.001) & (0.001) & (0.001) \\ 
  & & & & & & & & & & & & \\ 
 Constant & 0.00001 & 0.00000 & $-$0.00000 & 0.00000 & 0.00002 & 0.00001 & 0.00001 & 0.00001 & 0.00001 & $-$0.00000 & $-$0.0001 & 0.00001 \\ 
  & (0.001) & (0.001) & (0.001) & (0.001) & (0.001) & (0.001) & (0.001) & (0.001) & (0.001) & (0.001) & (0.001) & (0.001) \\ 
  & & & & & & & & & & & & \\ 
\hline \\[-1.8ex] 
Observations & 845,773 & 845,773 & 845,773 & 845,773 & 845,773 & 845,773 & 845,773 & 845,773 & 845,773 & 845,773 & 845,773 & 845,773 \\ 
R$^{2}$ & 0.013 & 0.005 & 0.101 & 0.003 & 0.031 & 0.035 & 0.001 & 0.004 & 0.002 & 0.013 & 0.002 & 0.002 \\ 
Adjusted R$^{2}$ & 0.013 & 0.005 & 0.101 & 0.003 & 0.031 & 0.035 & 0.001 & 0.004 & 0.002 & 0.013 & 0.002 & 0.002 \\ 
Residual Std. Error & 0.994 & 0.998 & 0.948 & 0.998 & 0.984 & 0.982 & 1.000 & 0.998 & 0.999 & 0.993 & 0.999 & 0.999 \\ 
\hline 
\hline \\[-1.8ex] 
\textit{Note:}  & \multicolumn{12}{r}{$^{*}$p$<$0.1; $^{**}$p$<$0.05; $^{***}$p$<$0.01} \\ 
\end{tabular} 
}
\end{table} 

\begin{table} \centering 
  \caption{Regression results for the models for different social, transportation, and mobility variables as a function of the weights of the latent behaviors and several demographic variables. See Supplementary Eqs. (\ref{eqSM:model1}) and (\ref{eqSM:model2}) . Models for Integration and Exploration are done at individual level. The rest of the models are done aggregating the variables over census tracts. Because of the normalization of weights in latent behaviors, we chose to exclude ``College``(Latent Behavior 4) in models for Integration and Exploration to prevent co-linearity.} 
  \label{tableSM:outcomes} 
\scalebox{0.86}{
\tabcolsep=0.2cm 
\begin{tabular}{@{\extracolsep{-5pt}}lcccccc} 
\\[-1.8ex]\hline 
\hline \\[-1.8ex] 
 & \multicolumn{6}{c}{\textit{Dependent variable $X_i$ or $X_\alpha$:}} \\ 
\cline{2-7} 
\\[-1.8ex] & Integration & Exploration & Long Commute & Distance travelled & Physical Activity & No obesity \\ 
\hline \\[-1.8ex] 
 Latent Behavior 1 & 0.141$^{***}$ & $-$0.208$^{***}$ & 0.125$^{***}$ & 0.146$^{***}$ & $-$0.008 & 0.013 \\ 
  & (0.002) & (0.002) & (0.014) & (0.008) & (0.009) & (0.009) \\ 
  & & & & & & \\[-1.8ex]  
 Latent Behavior 2 & $-$0.102$^{***}$ & 0.122$^{***}$ & $-$0.036$^{***}$ & $-$0.038$^{***}$ & $-$0.096$^{***}$ & $-$0.013$^{*}$ \\   
  & (0.001) & (0.001) & (0.012) & (0.006) & (0.008) & (0.007) \\ 
  & & & & & & \\[-1.8ex]  
 Latent Behavior 3 & $-$0.020$^{***}$ & 0.052$^{***}$ & 0.247$^{***}$ & $-$0.186$^{***}$ & $-$0.131$^{***}$ & $-$0.105$^{***}$ \\ 
  & (0.002) & (0.001) & (0.017) & (0.010) & (0.012) & (0.011) \\ 
  & & & & & & \\[-1.8ex]   
 Latent Behavior 4 &  &  & 0.006 & $-$0.014$^{**}$ & $-$0.078$^{***}$ & $-$0.079$^{***}$ \\ 
  &  &  & (0.011) & (0.006) & (0.007) & (0.007) \\ 
  & & & & & & \\[-1.8ex] 
 Latent Behavior 5 & 0.071$^{***}$ & 0.119$^{***}$ & 0.121$^{***}$ & 0.051$^{***}$ & 0.223$^{***}$ & 0.163$^{***}$ \\ 
  & (0.002) & (0.002) & (0.014) & (0.007) & (0.009) & (0.009) \\ 
  & & & & & & \\[-1.8ex] 
 Latent Behavior 6 & $-$0.155$^{***}$ & 0.214$^{***}$ & $-$0.076$^{***}$ & $-$0.118$^{***}$ & $-$0.086$^{***}$ & $-$0.121$^{***}$ \\ 
  & (0.001) & (0.001) & (0.014) & (0.007) & (0.009) & (0.009) \\ 
  & & & & & & \\[-1.8ex] 
 Latent Behavior 7 & 0.014$^{***}$ & 0.001 & 0.053$^{***}$ & $-$0.030$^{***}$ & $-$0.011$^{*}$ & $-$0.001 \\ 
  & (0.001) & (0.001) & (0.010) & (0.005) & (0.006) & (0.006) \\ 
  & & & & & & \\[-1.8ex] 
 Latent Behavior 8 & 0.043$^{***}$ & $-$0.024$^{***}$ & 0.095$^{***}$ & $-$0.014$^{**}$ & 0.001 & 0.010 \\ 
  & (0.001) & (0.001) & (0.011) & (0.006) & (0.007) & (0.007) \\ 
  & & & & & & \\[-1.8ex] 
 Latent Behavior 9 & 0.070$^{***}$ & $-$0.052$^{***}$ & 0.135$^{***}$ & $-$0.013$^{**}$ & $-$0.0004 & 0.005 \\ 
  & (0.002) & (0.002) & (0.010) & (0.006) & (0.007) & (0.006) \\ 
  & & & & & & \\[-1.8ex] 
 Latent Behavior 10 & $-$0.205$^{***}$ & 0.382$^{***}$ & 0.127$^{***}$ & 0.117$^{***}$ & $-$0.041$^{***}$ & $-$0.023$^{***}$ \\ 
  & (0.002) & (0.002) & (0.012) & (0.007) & (0.008) & (0.008) \\ 
  & & & & & & \\[-1.8ex] 
 Latent Behavior 11 & $-$0.020$^{***}$ & $-$0.016$^{***}$ & 0.125$^{***}$ & $-$0.050$^{***}$ & $-$0.046$^{***}$ & $-$0.043$^{***}$ \\
  & (0.002) & (0.002) & (0.011) & (0.006) & (0.007) & (0.007) \\ 
  & & & & & & \\[-1.8ex] 
 Latent Behavior 12 & 0.023$^{***}$ & $-$0.053$^{***}$ & 0.049$^{***}$ & $-$0.051$^{***}$ & $-$0.030$^{***}$ & $-$0.035$^{***}$ \\ 
  & (0.002) & (0.002) & (0.012) & (0.007) & (0.008) & (0.008) \\ 
  & & & & & & \\[-1.8ex] 
 Population Density & 0.006$^{***}$ & 0.019$^{***}$ & $-$0.082$^{***}$ & $-$0.028$^{***}$ & 0.024$^{***}$ & 0.007 \\ 
  & (0.001) & (0.001) & (0.016) & (0.009) & (0.008) & (0.008) \\ 
  & & & & & & \\[-1.8ex] 
 Median household & $-$0.103$^{***}$ & 0.070$^{***}$ & 0.0002 & 0.690$^{***}$ & $-$0.528$^{***}$ & $-$0.307$^{***}$ \\ 
 income & (0.001) & (0.001) & (0.012) & (0.007) & (0.008) & (0.008) \\ 
  & & & & & & \\[-1.8ex] 
 Fraction of users of & 0.054$^{***}$ & 0.054$^{***}$ & 0.158$^{***}$ & $-$0.069$^{***}$ & 0.029$^{**}$ & $-$0.037$^{***}$ \\ 
 public transp.  & (0.001) & (0.001) & (0.016) & (0.011) & (0.013) & (0.012) \\ 
  & & & & & & \\[-1.8ex] 
 Fraction of black & 0.090$^{***}$ & 0.001 & 0.081$^{***}$ & $-$0.093$^{***}$ & 0.062$^{***}$ & 0.377$^{***}$ \\ 
 population & (0.001) & (0.001) & (0.010) & (0.006) & (0.007) & (0.007) \\
  & & & & & & \\[-1.8ex] 
 Constant & 0.171$^{***}$ & $-$0.161$^{***}$ & $-$0.017 & $-$0.145$^{***}$ & $-$0.151$^{***}$ & $-$0.187$^{***}$ \\ 
  & (0.006) & (0.006) & (0.040) & (0.019) & (0.027) & (0.026) \\ 
  & & & & & & \\ 
\hline \\[-1.8ex] 
Observations & 845,773 & 845,773 & 7,023 & 13,179 & 5,710 & 5,710 \\ 
R$^{2}$ & 0.164 & 0.260 & 0.508 & 0.724 & 0.820 & 0.837 \\ 
Adjusted R$^{2}$ & 0.164 & 0.260 & 0.506 & 0.724 & 0.819 & 0.836 \\ 
Residual Std. Error & 0.914  & 0.861  & 0.703  & 0.526 & 0.425  & 0.405  \\ 
                    &  (df = 845747) &  (df = 845747) &  (df = 6998) &  (df = 13154) & (df = 5685) &  (df = 5685) \\ 

\hline 
\hline \\[-1.8ex] 
\textit{Note:}  & \multicolumn{6}{r}{$^{*}$p$<$0.1; $^{**}$p$<$0.05; $^{***}$p$<$0.01} \\ 
\end{tabular} 
}
\end{table}

\begin{table} \centering 
  \caption{Regression results for the models for different social, transportation, and mobility variables as a function only of demographic variables. See Supplementary Eqs. (\ref{eqSM:model1}) and (\ref{eqSM:model2}), and compare with Table \ref{tableSM:outcomes}.} 
  \label{tableSM:outcomesdemo} 
  \scalebox{0.8}{
\begin{tabular}{@{\extracolsep{5pt}}lcccccc} 
\\[-1.8ex]\hline 
\hline \\[-1.8ex] 
 & \multicolumn{6}{c}{\textit{Dependent variable $X_i$ or $X_\alpha$:}} \\ 
\cline{2-7} 
\\[-1.8ex] & Integration & Exploration & Long commute & Distance travelled & Physical Activity & No obesity\\ 
\hline \\[-1.8ex] 
 Population Density & $-$0.0002 & 0.021$^{***}$ & $-$0.097$^{***}$ & $-$0.118$^{***}$ & $-$0.031$^{***}$ & $-$0.048$^{***}$ \\ 
  & (0.001) & (0.001) & (0.016) & (0.011) & (0.010) & (0.009) \\ 
  & & & & & & \\ 
 Median household & $-$0.135$^{***}$ & 0.076$^{***}$ & $-$0.072$^{***}$ & 0.565$^{***}$ & $-$0.728$^{***}$ & $-$0.474$^{***}$ \\ 
 income & (0.001) & (0.001) & (0.010) & (0.007) & (0.008) & (0.007) \\ 
  & & & & & & \\ 
 Fraction of users of & 0.049$^{***}$ & 0.049$^{***}$ & 0.105$^{***}$  & $-$0.372$^{***}$ & $-$0.136$^{***}$ & $-$0.185$^{***}$ \\
 public transp. & (0.002) & (0.002) & (0.016)  & (0.010) & (0.013) & (0.012) \\ 
  & & & & & & \\ 
 Fraction of black & 0.108$^{***}$ & $-$0.005$^{***}$ & 0.154$^{***}$ & $-$0.057$^{***}$ & 0.174$^{***}$ & 0.461$^{***}$ \\ 
 population & (0.001) & (0.001) & (0.010) & (0.006) & (0.008) & (0.007) \\ 
  & & & & & & \\ 
 Constant & 0.139$^{***}$ & $-$0.120$^{***}$ & $-$0.161$^{***}$ & $-$0.086$^{***}$ & $-$0.250$^{***}$ & $-$0.264$^{***}$ \\ 
  & (0.006) & (0.006) & (0.040) & (0.021) & (0.031) & (0.029) \\ 
  & & & & & & \\ 
\hline \\[-1.8ex] 
Observations & 845,773 & 845,773 & 7,023 & 13,179 & 5,710 & 5,710 \\ 
R$^{2}$ & 0.059 & 0.025 & 0.454 & 0.625 & 0.739 & 0.781 \\ 
Adjusted R$^{2}$ & 0.059 & 0.025 & 0.453 & 0.625 & 0.739 & 0.781 \\ 
Residual Std. Error & 0.970  & 0.987  & 0.740  & 0.613  & 0.511  & 0.468  \\ 
 & (df = 845758) &  (df = 845758) &  (df = 7010) &  (df = 13166) &  (df = 5697) &  (df = 5697) \\ 
\hline 
\hline \\[-1.8ex] 
\textit{Note:}  & \multicolumn{6}{r}{$^{*}$p$<$0.1; $^{**}$p$<$0.05; $^{***}$p$<$0.01} \\ 
\end{tabular} 
}
\end{table}

\end{document}